\documentclass[a4paper,twocolumn,amsmath,pra,showpacs,amssymb]{revtex4-1} 
\usepackage[T1]{fontenc}
\usepackage[latin9]{inputenc}
\usepackage{times}
\usepackage{color}
\usepackage{xspace}
\usepackage{amssymb,amsmath}
\usepackage{amsbsy}
\usepackage{graphicx}
\usepackage{multirow}
\usepackage{rotating}
\usepackage{microtype}
\usepackage{tabularx}
\usepackage{epstopdf}
\usepackage{float}
\usepackage{mathrsfs}
\usepackage{hyperref}

\usepackage{calligra}
\DeclareMathAlphabet{\mathcalligra}{T1}{calligra}{m}{n}
\DeclareFontShape{T1}{calligra}{m}{n}{<->s*[2.2]callig15}{}

\def\be{\begin{equation}}
\def\ee{\end{equation}}
\def\e#1{\label{#1}\end{equation}}
\def\bea{\begin{eqnarray}}
\def\eea{\end{eqnarray}}
\def\ea#1{\label{#1}\end{eqnarray}}

\def\bem#1{\begin{mathletters}\label{#1}}
\def\eml{\end{mathletters}}

\def\ket#1{{|#1\rangle}}
\def\bra#1{{\langle#1|}}

\def\4#1{{\boldsymbol{#1}}}
\def\8#1{{\widetilde{#1}}}
\def\bse{\begin{subequations}}
\def\ese{\end{subequations}}
\def\nn{\nonumber}

\def\Rb87{$^{87}\text{Rb}$}
\def\0{\ket{0}}
\def\1{\ket{1}}

\begin{document}
\title{Robust Rydberg interaction gates with adiabatic passage}
\author{D. D. Bhaktavatsala Rao}
\author{Klaus M{\o}lmer}
\affiliation{%
Department of Physics and Astronomy, Aarhus University, Ny Munkegade 120, DK-8000, Aarhus C, Denmark. \\
}%
\date{\today}
\begin{abstract}
We show that with adiabatic passage, one can reliably drive two-photon optical transitions between the ground states and interacting Rydberg states in a pair of atoms. For finite Rydberg interaction strengths a new adiabatic pathway towards the doubly Rydberg excited state is identified when a constant detuning is applied with respect to an intermediate optically excited level. The Rydberg interaction among the excited atoms provides a phase that may be used to implement quantum gate operations on atomic ground state qubits.
 \end{abstract}
\maketitle

The strong long range interaction between atoms excited to high-lying Rydberg states make cold atoms, trapped in tweezer arrays or optical lattices promising candidates for the implementation of different quantum computing protocols as well as for the study of a variety of complex many-body and light-matter problems \cite{revmod}. In the seminal proposal for quantum computing with neutral atoms Jaksch et al \cite{jaks} have shown that controlled phase operations between two neutral atoms can be performed either when the atoms experience very strong or very weak dipole-dipole (Rydberg) interactions. In the former case i.e., in the Rydberg blockade regime one excited atom causes a sufficiently large energy shifts of Rydberg states in a neighboring atom to effectively detune it away from resonance and fully block its excitation by a laser field. Controlled two-qubit gates and generation of entanglement have been demonstrated experimentally in the blockade regime \cite{exp1,exp2,exp3}. In the weak interaction regime both atoms can be excited to Rydberg states and the resulting doubly excited state attains a phase due to the weak Rydberg interaction. The two-atom gate proposals have been followed by a variety of schemes for fast quantum gates with atomic ensembles \cite{luk, saff1,saff2,zheng}, entangled state preparation \cite{klaus1}, quantum algorithms \cite{chen,klaus2}, quantum simulators \cite{weimer}, and efficient quantum repeaters \cite{exp4}. Other schemes using, e.g., the interaction in an anti-blockade regime \cite{pohl}, and in conjunction with strong dissipation \cite{durga,saffman}, have been proposed.

In this work, we shall propose and investigate a protocol that applies to the case of Rydberg interaction strengths that are too weak to yield the blockade mechanism, yet too large to be ignored when the atoms are excited with resonant laser fields. This intermediate regime applies to atoms beyond nearest neighbor separation in multi-atom architectures \cite{lattice}, and reliable two-qubit gates between such atoms are a prerequisite for effective implementation of algorithms on multi-qubit registers. The key ingredient in interaction based controlled gates is the possibility to reliably excite two atoms to the Rydberg states, and we shall apply adiabatic passage \cite{rap}, which is a widely used technique for robust transfer of population between different states in atomic and molecular systems. In ensembles, it has thus been proposed as a means to suppress the dependence of the single Rydberg atom excitation rate on the number of atoms and to achieve deterministic excitation for a range of different ensemble sizes \cite{beterov,david}.

We shall introduce the adiabatic dark states of coherently driven three-level atoms, and discuss how they are affected by the interaction between the excited states. We shall then determine adiabatic eigenstates of the two-atom Hamiltonian, and show how a constant detuning of the driving fields can be used to obtain a new adiabatic pathway towards the doubly Rydberg excited state. Finally, we include dissipation and atomic motion to estimate gate fidelities and optimal parameters for the gate operation.

\begin{figure}
\includegraphics[width=100mm]{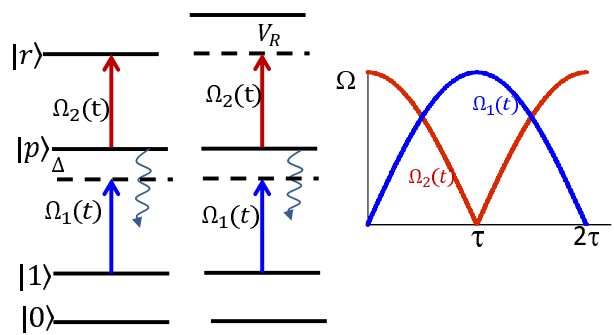}
\label{setup1}

\caption{(Color online) Energy-level diagram  of atoms driven by two-photon laser transitions with Rabi frequencies $\Omega_1(t),\Omega_2(t)$ between ground $|1\rangle$ and Rydberg states $|r\rangle$. The pair of Rydberg states experience the interaction $V_R$. The right hand side of the figure shows the counter-intuitive pulse sequence for the Rabi fields $\Omega_1(t), \Omega_2(t)$, driving the atoms from $|1\rangle$ to $|r\rangle$ and back. This process yields a phase gate on the ground qubit states $\ket{0}$, $\ket{1}$.}
\end{figure}

We consider two atoms with two ground hyperfine states, labeled $\ket{0}$ and $\ket{1}$, and a Rydberg state $\ket{r}$, which can be excited via the intermediate state $\ket{p}$, see Fig. 1. The atoms experience a non-zero energy shift $V_R$, when both atoms occupy the Rydberg state $\ket{r}$. The transition Rabi frequencies between states $\ket{1}$ and $\ket{p}$ and between $\ket{p}$ and $\ket{r}$  are denoted $\Omega_1(t)$ and $\Omega_2(t)$, respectively, and a detuning $\Delta$ may be applied with respect to the intermediate level $\ket{p}$. The Rydberg and optically excited states decay by spontaneous emission of radiation, and we assume that the Rydberg state lifetime is much longer than the optical state lifetime $\gamma_r << \gamma_p$. The total Hamiltonian describing the interacting Rydberg atoms is
\be
{H}(t) = H_1\otimes\mathcal{I}+\mathcal{I}\otimes H_2 + V_R\ket{rr}\bra{rr},
\ee
with single atom Hamiltonian operators ($\hbar = 1$),
\bea
\label{ham}
H_{j} = \Delta\ket{p}_{jj}\bra{p} + \Omega_1(t)\ket{1}_{jj}\bra{p}+\Omega_2(t)\ket{p}_{jj}\bra{r}+h.c.,
\eea
($j=1,2$). For concreteness, we consider the following time-dependent Rabi frequencies ( $0 \le t \le 2  \tau$)
\be
\Omega_{1}(t)=\Omega\sin\left(\frac{\pi}{2\tau}t\right)~ \Omega_{2}(t)=\Omega|\cos\left(\frac{\pi}{2\tau}t\right)|.
\ee
When $\Delta=V_R=0$, this choice leads to adiabatic eigenstates with equidistant energies, separated by $\Omega$.

The atoms initially populate the ground atomic states, and when $V_R=0$, the $|1\rangle$ component of both atoms  adiabatically follow the dark state (zero eigenvalue state at any time $t$)
\be
\ket{D}\equiv \cos\theta \ket{1}-\sin\theta\ket{r},~~\tan\theta=\Omega_1(t)/\Omega_2(t).
\ee
Hence, as long as the adiabatic condition, $\Omega\tau \gg 1$ is fulfilled, i.e., $\tau$ is chosen long enough,
all the population in the product state $\ket{11}$ is transferred to the doubly excited Rydberg state $\ket{rr}$ and back without populating the intermediate level $\ket{p}$ of either of the atoms.

In the limit of strong interaction i.e., $V_R > \Omega \gg 1/\tau$, the dark state dynamics is destroyed, resulting in optical excitation and spontaneous decay \cite{david}. At the end of the pulse sequence (3) there is an equal probability to find either of the atoms in the Rydberg state, while the other atom may have experienced several spontaneous emission events and the coherence and entanglement between the atoms is lost.

In the opposite limit of weak interaction, i.e., $V_R \ll \Omega$, we can treat the interaction term in (1) as a perturbation. To lowest order, the adiabatic eigenstates attain the same form as in the absence of $V_R$, but it causes an energy shift that we compute as the expectation value of the interaction part of the Hamiltonian in the time dependent dark state $|DD\rangle$. This merely yields the constant $V_R$ scaled by the $|rr\rangle$ population of $|DD\rangle$, $P_{rr}(t) = \sin^4\theta$, and the dark state thus accumulates the corresponding time dependent phase
\be
\phi(t) = V_R\int^t_0 P_{rr}(t') dt'.
\ee
In a single cycle of the Rabi frequency evolution (3), sketched in Fig. 1, the atoms in state $|1\rangle$ are excited to state $|r\rangle$ and back, and the product state $\ket{11}$ attains a phase $3\tau V_R/8$ after time $2\tau$, while all other states $\ket{10},\ket{01},\ket{00}$ do not get this phase as they do not populate the doubly excited state $\ket{rr}$. Thus one achieves a controlled phase gate, which in the ground state basis $\ket{11}, ~\ket{10}, ~\ket{01}, ~\ket{00}$ can be written as
\be
U_\phi =  \left[ \begin{array}{cccc} {\rm e}^{i\phi} & 0 & 0 & 0 \\ 0 & 1 & 0 & 0 \\0 & 0 & 1 & 0  \\0 & 0 & 0 & 1  \end{array} \right],
\ee
and for $\tau=\frac{8\pi}{3V_R}$ we achieve the maximally entangling $\pi$-phase gate for two atoms. Since $V_R \ll \Omega$, this gate is slow and thus subject to Rydberg state dephasing and decay.

We are now interested in the intermediate regime, $V_R \sim \Omega$. The interaction term thus constitutes a non-negligible detuning for the transition to the state $|rr\rangle$, and in case of constant amplitude resonant excitation, the atoms will populate undesired state components, including the short lived intermediate state, resulting in a loss of gate fidelity. We will now analyze the dynamics of the system by first solving the unitary dynamics generated by the Hamiltonian given in (1). This will be followed by a master equation analysis of the effects of dissipation.

\begin{figure}
\begin{center}
\includegraphics[width=90mm]{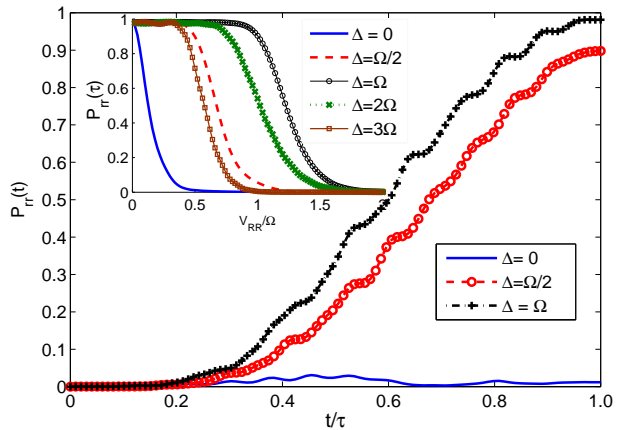}
\end{center}
\label{setup1}
\caption{(Color online) The population of the doubly excited Rydberg state $P_{rr}(\tau)$ is shown as a function of time for fixed $V_R = \Omega/2$ and pulse length $\tau = 0.25 \mu s$. The different curves correspond to different values of the intermediate state detuning $\Delta$. In the inset we show for different values of $\Delta$, how $P_{rr}(t=\tau)$ varies as a function of the Rydberg interaction strength $V_R$. Both atoms are initially in their ground states, $\ket{\psi(0)} = \ket{11}$, and all calculations assume $\Omega/2\pi = 50$ MHz.}
\end{figure}

In Fig. 2 we show the $|rr\rangle$ population, $P_{rr}(t)$ as a function of time $t \in [0,\tau]$ for $V_R = \Omega/2$. The lower, blue curve in the figure shows the results for a vanishing detuning, $\Delta=0$, with respect to the intermediate state, and we see that the excitation of two Rydberg atoms is almost completely suppressed. The reason for this can be understood by a closer examination of the adiabatic eigenstates of the Hamiltonian. $|11\rangle$ and $|rr\rangle$ are, indeed, eigenstates at times $t=0$ and $t=\tau$, respectively, but, they are not adiabatically connected: For $\Delta=0$, the former state evolves along the interacting two-atom dark state \cite{klaus1}
\bea \label{dark2}
\ket{\psi(\theta)} &=& \frac{1}{\sqrt{\cos^4\theta + 2\sin^4\theta}}\big[(\cos^2\theta-\sin^2\theta)\ket{11} \nn \\
&-&\cos\theta\sin\theta(\ket{1r}+\ket{r1})+\sin^2\theta\ket{pp}\big].
\eea
Adiabatic following of this state does not populate the doubly excited Rydberg state at $t=\tau~(\theta=\pi/2)$, but it populates the $|p\rangle$-state and suffers from spontaneous emission. At $t=\tau$, the state $|rr\rangle$ is instead one of the other eigenstates of the time dependent Hamiltonian, which adiabatically correlates with a state that populates the doubly excited Rydberg state with a probability of $0.5$ for times $t=0, ~2\tau$. In Fig. 3a, for the same parameters as in Fig. 2, we show the eigenenergies of the time dependent Hamiltonian. The middle (blue) and the upper (red) dashed curves show the energies for the two atom dark state and for the state attaining $|rr\rangle$ at $t=\tau$, respectively. The curves approach each other, and the small population of $|rr\rangle$ in Fig. 2 (lower, blue curve) can be understood as due to non-adiabatic transitions between the adiabatic eigenstates.

\begin{figure}
\begin{center}
\includegraphics[width=90mm]{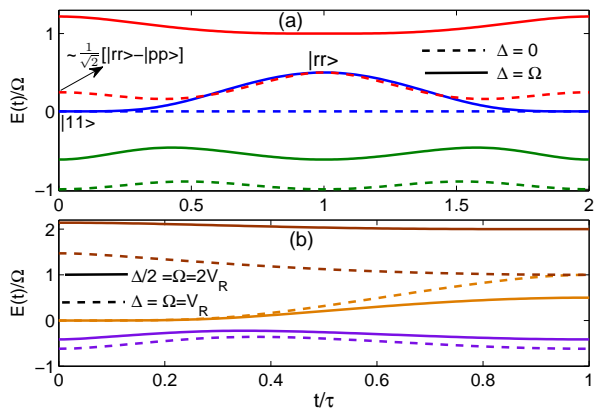}
\end{center}
\label{setup1}
\caption{(Color online) Three energy eigenvalues are shown for the time dependent Hamiltonian (1); (a) for vanishing $\Delta$ (solid lines) and for $\Delta=\Omega=2V_R$ (dashed lines) and (b) for $\Delta/2 = \Omega = 2V_R$ (solid lines) and for $\Delta = \Omega = V_R$ (dashed lines).The values of $\Omega$ and $\tau$ are the same as those used in Fig. 1}
\end{figure}

The picture changes completely, when we apply a finite detuning from the middle level. This breaks the degeneracy (in the rotating frame) of the states in (\ref{dark2}), and if the detuning is chosen to be comparable to $V_R$ and $\Omega$, new adiabatic eigenstates are formed with significantly modified character. The solid lines in Fig. 3a, show the eigenenergies of the time dependent Hamiltonian for the case $\Delta=\Omega=2V_R$ and now, indeed, the state $|11\rangle$ \textit{is} adiabatically connected to $|rr\rangle$ along the middle (blue) solid line.

The time dependent population of the the state $|rr\rangle$ during the time evolution with $\Delta=\Omega=2V_R$ is shown as the upper, black curve in Fig. 2, and indeed, it reaches a value near unity at $t=\tau$. Proceeding and evolving the system adiabatically back to the state $|11\rangle$ at $t=2\tau$, the Rydberg interaction leads to an accumulated phase, proportional to the area between the middle, blue solid and dashed curves in Fig. 3a, and the corresponding phase gate on the ground qubit state manifold has been implemented.

Let us now address the conditions for high fidelity phase gate performance with $\Delta,\ \Omega$, and $V_R$ of comparable magnitude.  While a small value of $V_R$ clearly permits high population transfer to $|rr\rangle$, our request for a finite $V_R$ to yield the desired phase evolution faster is fulfilled if we take $\Omega \sim V_R$ and $\Delta \sim \Omega$. The detuning should not be taken much larger, since that would cause non-adiabatic transfer towards the low energy eigenstate (lower (violet) solid line in Fig. 3b), while making $V_R$ much larger than $\Omega$ will cause non-adiabatic transfer towards high energy eigenstate (upper (brown) dashed line in Fig. 3b).

Our calculations have shown that it is possible to rapidly transfer atoms from $|11\rangle$ to $|rr\rangle$ and back. To ensure adiabaticity, the total time must be long enough, $\Omega \tau \gtrsim 1$. This constraint is similar to the conventional single atom adiabaticity condition, which must anyway be fulfilled such that the qubit product states $|01(10)\rangle$ evolve adiabatically along $|0D(D0)\rangle$. The phase  $\sim V_R \tau$ accumulated on $|11\rangle$ may be readily adjusted to the value $\pi$, needed for a maximally entangling phase gate.


\begin{figure}
\begin{center}
\includegraphics[width=90mm]{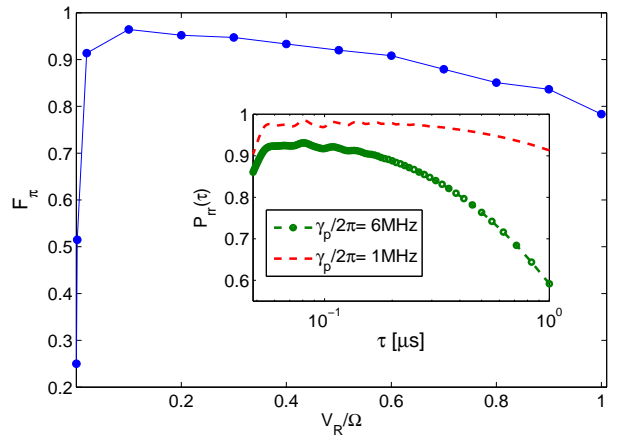}
\end{center}
\label{setup1}
\caption{(Color online) The optimal final state fidelity for the $\pi$-phase gate, $F_\pi \equiv Tr[U_\pi \rho(0) U_\pi^\dagger\rho(2\tau)]$, applied to atoms initialized in $\rho(0) = \ket{\psi}\bra{\psi}$, where $\ket{\psi} = \frac{1}{\sqrt{2}}[\ket{1}+\ket{0}]$, is shown as a function of the Rydberg interaction strength $V_R$. The decay rates applied in the simulations are  $\gamma_p/2\pi = 6$MHz, $\gamma_r/2\pi = 1$kHz, $\gamma_{rd}/2\pi = 10$kHz. In the inset we show the double excitation probability at the end of half cycle (staring in $|11\rangle$), $P_{rr}(\tau)$, as a function of $\tau$ for $\gamma_p/2\pi = 1.01$MHz (red dashed-line) and $\gamma_p/2\pi = 6$MHz (green circles). The strength of the interaction and the single photon Rabi frequency are chosen to be $V_R/2\pi = 25$MHz and $\Omega/2\pi = 50$MHz.}
\end{figure}

Due to the Rydberg interaction and the detuning the adiabatic state leading to the double excitation of the Rydberg state will have a nonzero population in $\ket{p}$ which is subject to the effect of spontaneous emission and decay, and the Rydberg state is also subject to decoherence and decay. To evaluate the dynamics in the presence of these dissipative effects we therefore solve the master equation
\be
\partial_t \rho = i[\mathcal{H}_{eff},\rho] + \sum_{j,k} \mathcal{C}^{{(j)}^\dagger}_k\rho\mathcal{C}^{(j)}_k,
\ee
where $\mathcal{H}_{eff} = \mathcal{H}-\frac{i}{2}\sum_{j,k}\mathcal{C}^{{(j)}^\dagger}_k\mathcal{C}^{(j)}_k$, and
$\mathcal{C}^{(j)}_k$ are Lindblad operators, which describe the decay processes of the $j-th$ atom by spontaneous emission of light, $\mathcal{C}^{(j)}_0 = \sqrt{\gamma_0}\ket{0}_{jj}\bra{p}, ~\mathcal{C}^{(j)}_1 = \sqrt{\gamma_1}\ket{1}_{jj}\bra{p}$ ($\gamma_p=\gamma_0 +\gamma_1$). Rydberg state decay is described by similar terms, \textit{e.g.}, $\mathcal{C}^{(j)}_r = \sqrt{\gamma_r}\ket{p}_{jj}\bra{r}$, but it is a much slower process, and we also consider the dephasing of the Rydberg state $\mathcal{C}^{(j)}_{rd} = \sqrt{\gamma_{rd}}(\mathcal{I}-2\ket{r}\bra{r})$.

In Fig. 4, we show results, where we have solved the master equation for systems starting in a product state $\frac{1}{2}(|0\rangle+|1\rangle)(|0\rangle+|1\rangle)$.  We applied $\Delta=\Omega= 2\pi\cdot50$MHz, and for different values of $V_R/\Omega$ we adjusted $\tau$ to obtain the desired $\pi$ phase shift. The curve in the figure shows the final state fidelity when the decay rates are taken as $\gamma_p/2\pi = 6$MHz, $\gamma_r/2\pi = 1$kHz, $\gamma_{rd}/2\pi = 10$kHz, corresponding to realistic parameters for Rb atoms. The fidelity is deteriorated for small values of $V_R$ due to dephasing and decay of the Rydberg states (as $\tau \sim 1/V_R$), while for large values of $V_R$  non-adiabatic processes are important (as $\tau \sim 1/\Omega$). In the inset we show how the population of the state $|rr\rangle$ at the time $\tau$ varies as a function of the duration of the adiabatic process. Both curves assume $\Omega/2\pi =50$ MHz, and $V_R=\Omega/2$. The upper curve is obtained with a decay rate of $\gamma_p/2\pi= 1.01$MHz, while the lower curve is obtained with $\gamma_p/2\pi=6$MHz, corresponding to Cs and Rb atoms respectively. The state transfer is better than 90$\%$ for both sets of parameters, and inspection of the parameters show that a duration as short as $\tau= 0.25\mu s$ leads to the desired $\pi$ phase shift and a phase gate with acceptable fidelity.

Atoms excited to the Rydberg state during the gate experience a force that excites and entangles their motional states with the qubit degree of freedom \cite{igor} and thus decreases the fidelity of the gate. We now estimate this loss of fidelity for atoms separated by a distance $r$, initialized in the ground states of tight harmonic traps with trapping frequency $\omega_0$ and a  position uncertainty $r_0$. For the $C_6/r^6$ van der Waals interaction, the atoms in $|rr\rangle$ experience a force $\sim\frac{6}{r}V_R$, and perturbation theory yields a motional excited state amplitude, $\sim\frac{6r_0}{r}\frac{V_R}{\omega_0}(1-{\rm e}^{-i\omega_0\tau})$. During the gate operation, the average population in the $|rr\rangle$ is on order of $1/2$ and thus reduces the error due to motional entanglement by that factor. For ${}^{87}$Rb atoms separated by a distance $\sim 11.73 \mu$m and excited to the high lying Rydberg state $(n=97$) \cite{mark}, the Rydberg interaction energy yields $V_R/2\pi = 4$MHz. With the single photon Rabi frequency $\Omega/2\pi = 50$MHz, the duration of the phase gate is $2\tau \sim 0.5\mu$s, and if the atoms are held in optical tweezer traps with $\omega_0/2\pi = 36$kHz and $r_0 = 390$nm, the probability of excitation to the first excited motional states during the gate time $2\tau$ is then only $\sim 0.022$. Using instead an optical lattice with atoms located at selected lattice sites, the motional frequency is much higher and the ground state width is an order of magnitude smaller, the effects of motional excitation may be ignored.

In conclusion, we have shown that atoms with intermediate value Rydberg interaction strengths, incompatible with the blockade mechanism, but too strong to be dealt with by perturbation theory, can be excited reliably by adiabatic passage and accumulate a well controlled phase, useful for entangling quantum gates.
The result may be important for quantum computing architectures, where some but not all atoms are within the blockade radius distance of each other, and where different variants of interaction gates may become useful. The interaction strength in our numerical example is typical for highly excited atoms at intermediate to large separation, and also for closer atoms excited to lower lying Rydberg states. By comparison, it is indeed possible to operate a blockade gate if the Rabi coupling frequency is kept well below the value of $V_R$. On resonance with the intermediate level that would, however, be subject to decay and a significantly lower fidelity, while for a large detuning $\Delta$, the requirements on laser coupling strengths become severe. Our simulations already assumed a Rabi frequency stronger than $V_R$, and we have also investigated the simple interaction gate for the same parameters, i.e., the direct resonant excitation to $|rr\rangle$ with the strong fields. This process, however, also suffers from fidelity loss due to decay from the intermediate optically excited state. The rapid adiabatic process in comparison seems to make optimum use of the robust near-resonant state transfer and the (almost) dark state character of the intermediate states.

\begin{acknowledgements}
        This work was supported by the project MALICIA under FET-Open grant number $265522$, and the IARPA MQCO program. The authors are grateful to Mark Saffman, David Petrosyan, Malte C. Tichy, Pinja Haikka, Christian K. Andersen, for discussions and useful comments on the manuscript.
\end{acknowledgements}

\end{document}